\documentclass[floatfix,amssymb,prl,twocolumn,superscriptaddress,preprintnumbers,nofootinbib, aps]{revtex4-1}
		
\usepackage{amssymb,amsmath,verbatim,mathtools,needspace,enumitem,etoolbox,graphicx,physics,microtype,afterpage,bm,soul}
\usepackage[dvipsnames]{xcolor}
\definecolor{linkcolor}{rgb}{0.0,0.3,0.5}
\usepackage[unicode, colorlinks=true, linkcolor=linkcolor, citecolor=linkcolor, filecolor=linkcolor,urlcolor=linkcolor, pdfusetitle]{hyperref}
\usepackage{orcidlink}
\usepackage[all]{hypcap}
\usepackage[T1]{fontenc}
\usepackage[utf8]{inputenc}
\usepackage{tabularx}
\usepackage{float}
\usepackage{nicefrac}
\usepackage{placeins}
\usepackage{lmodern}
\usepackage{ragged2e}
\allowdisplaybreaks
\usepackage{tikz}
\usepackage{color}
\usepackage{framed}
\usepackage{hyperref}
\hypersetup{colorlinks, citecolor=rossos, linkcolor=black, urlcolor=rossos}
\definecolor{rossos}{cmyk}{0,1,1,0.55}
\definecolor{bluscuro}{rgb}{0.15, 0.2, .85}
\definecolor{bluchiaro}{cmyk}{1,.3,0.,0.1}
\definecolor{ForestGreen}{rgb}{0.13, 0.55, 0.13}
\definecolor{darkblue}{rgb}{0,0, 1.39}

\newcommand{\be}{\begin{equation}}
\newcommand{\ee}{\end{equation}}
\renewcommand{\d}{{\rm d}}

\newcommand{\PBH}{\text{\tiny PBH}}
\newcommand{\WD}{\text{WD}}

\usepackage{pifont}

\def\lsim{\mathrel{\rlap{\lower4pt\hbox{\hskip0.5pt$\sim$}}
    \raise1pt\hbox{$<$}}}         
\def\gsim{\mathrel{\rlap{\lower4pt\hbox{\hskip0.5pt$\sim$}}
    \raise1pt\hbox{$>$}}}         

\newcommand{\jhu}{William H.\ Miller III Department of Physics and Astronomy, Johns Hopkins University, \\ 3400 North Charles Street, Baltimore, Maryland, 21218, USA}

\newcommand{\utica}{Department of Mathematics and Physics, SUNY Polytechnic Institute, 100 Seymour Road, Utica, NY 13502, USA}

\newcommand{\Kavli}{Kavli Institute for the Physics and Mathematics of the Universe (WPI), The University of Tokyo Institutes for Advanced Study, The University of Tokyo, Kashiwa, Chiba 277-8583, Japan}

\newcommand{\UCLA}{Department of Physics and Astronomy, University of California, Los Angeles California, 90095-1547, USA}

\begin{document}

\title{Gravitational-Wave Fingerprints of White Dwarfs Ignited by Primordial Black Holes}

\author{Loris Del Grosso\orcidlink{0000-0002-6722-4629}}
\email{ldelgro1@jh.edu}
\affiliation{\jhu}

\author{Alexander Kusenko\orcidlink{0000-0002-8619-1260}}
\email{kusenko@ucla.edu}
\affiliation{\UCLA}
\affiliation{\Kavli}

\author{Shing-Chi Leung\orcidlink{0000-0002-4972-3803}}
\email{leungs@sunypoly.edu}
\affiliation{\utica}
\affiliation{\Kavli}

\author{Surjeet Rajendran\orcidlink{0000-0001-9915-3573}}
\email{srajend4@jhu.edu}
\affiliation{\jhu}
\preprint{ET-0874A-26}
\begin{abstract}
\medskip
\noindent
%
An asteroid-mass primordial black hole (PBH) transiting a white dwarf can ignite a thermonuclear runaway, leading to a Type Ia supernova. We show that gravitational waves (GWs) can distinguish this channel from the standard astrophysical ones: a double-degenerate merger is preceded by a milli- to decihertz inspiral, while a single-degenerate explosion, ignited near the center, leaves only a weak GW memory. Using hydrodynamical simulations, we find that a white dwarf ignited off-center by an asteroid-mass PBH produces a GW memory $\sim$ ten times larger than in the single-degenerate channel, detectable by DECIGO out to $\simeq 280\,{\rm kpc}$, which covers the Galactic halo and its satellite galaxies. Should such an event be observed, it would point to asteroid-mass PBHs as constituting all of the dark matter. By contrast, we find that the stochastic GW background is undetectable, even assuming that the entire cosmological population of Type Ia supernovae is PBH-triggered. 
\end{abstract}
\maketitle

\vspace{0.1cm}
\noindent{{\bf{\em Introduction.}}}
Type~Ia supernovae (SNe~Ia) are thermonuclear explosions of carbon--oxygen white dwarfs (WDs), yet the nature of their progenitors and the mechanism that triggers the explosion are still debated~\cite{Maoz2014}. Progenitor scenarios are commonly classified by the number of degenerate stars (i.e., WDs) in the system. In the single-degenerate (SD) channel, a WD accretes matter from a non-degenerate companion in a binary, such as a red giant or a main-sequence star, until it approaches the Chandrasekhar mass ($M_{\rm Ch} \simeq 1.4\,M_\odot$) and ignites~\cite{Whelan1973, Nomoto1984}. In the double-degenerate (DD) channel, two WDs merge after an inspiral driven by gravitational-wave (GW) emission~\cite{Iben1984, Webbink1984, Pakmor2010, Pakmor2012}. In this case, the explosion may be triggered when the less massive WD is tidally disrupted and accreted by the more massive one, or earlier, through a double detonation initiated in a helium-rich surface layer~\cite{Fink2010, Shen2018, Sim:2010kc, Pollin:2024cez, Moran-Fraile:2023xms}. Electromagnetic (EM) searches for signatures of the companion have so far been inconclusive~\cite{Kasen2010, Gilfanov2010, Schaefer2012, Maoz2014}.
\begin{figure}[t]
  \centering
  \includegraphics[width=\columnwidth]{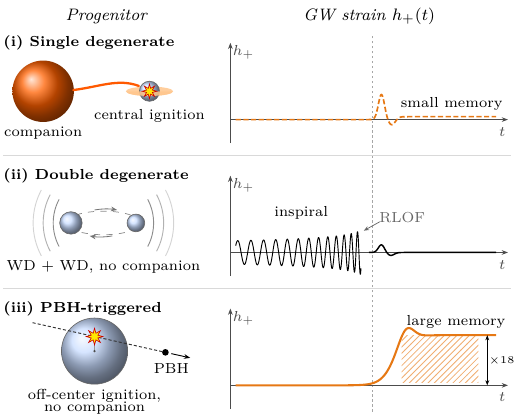}
  \caption{Gravitational-wave fingerprints of SN~Ia progenitors 
  (schematic, not to scale). (i) Single degenerate: a non-degenerate companion and
  near-central ignition leave a sharp GW burst but a small memory. (ii) Double degenerate:
  the explosion is preceded by a milli- to decihertz inspiral, truncated at Roche-lobe
  overflow (RLOF); the memory is model dependent. (iii) PBH-triggered: an isolated WD
  ignited off-center ($R_{\rm ign}\simeq0.44\,R_{\rm WD}$) leaves a memory $\simeq18$
  times larger than in (i), $d_{\rm L}\Delta h\simeq22\,{\rm cm}$ (hatched: memory gap).}
  \label{fig:smokinggun}
\end{figure}
A third possibility is that an isolated WD is ignited by a transiting primordial black hole (PBH)~\cite{Graham2015, MC19, Smirnov:2022zip, Leung:2025enj, Steigerwald:2021vgi, Steigerwald:2025pmu, Ganguly:2025ffb}  in the asteroid-mass window, $10^{-16}\lesssim M_\PBH/M_\odot \lesssim 10^{-11}$, where PBHs can account for all of the dark matter (DM)~\cite{MC19, Carr2020, Green2021, Fuller:2017uyd, Tran:2023jci, DeLuca:2019llr, DeLuca:2021hde, Magee:2018opb}. As a PBH of mass $M_\PBH$ crosses the star with a velocity $v_\PBH$ comparable to the stellar escape velocity $v_{\rm esc}\sim 10^{-2}\,c$, it heats the surrounding medium through dynamical friction~\cite{Cantiello:2026eur}: nuclei at a distance $R$ from its trajectory experience an acceleration $\sim G M_\PBH/R^2$ over a time $\sim R/v_\PBH$, and thus acquire a velocity $\sim G M_\PBH/(R\, v_\PBH)$. Requiring this velocity to reach the thermal velocity $\sqrt{\Delta T/m_N}$, the PBH raises the temperature by $\Delta T$ within a cylinder of radius
\begin{align}\label{eq:cylinder}
	R &\approx 2 \times 10^{-3} \, {\rm cm} \left(\frac{C_{\mathcal{I}}}{10}\right)
    \left(\frac{m_N/\Delta T}{10^{3}}\right)^{\nicefrac{1}{2}} \nonumber \\
    &\quad \times \left(\frac{M_\text{\tiny PBH}}{10^{\,21}\,{\rm g}} \right)
    \left(\frac{v_\text{\tiny PBH}}{10^{-2}}\right)^{-1} \,,
\end{align}
where $m_N$ is the nucleon mass and the factor of $C_{\mathcal{I}}$ accounts for the Coulomb logarithm~\cite{Smirnov:2022zip}. Since degenerate matter cannot expand and cool, once a region larger than the trigger size $\lambda_{\rm min}$ (computed numerically in Ref.~\cite{Timmes1992} as a function of the WD density) is heated to $\Delta T\sim 1\,{\rm MeV}$, the nuclear energy generation rate exceeds the rate of thermal diffusion; a thermonuclear runaway then develops and disrupts the star~\cite{Timmes1992, Graham2015, MC19}. 
Interestingly, if such PBHs constitute all of the DM, they would trigger up to a few explosions per century in the Milky Way, comparable to, or slightly higher than, the observed SN~Ia rate~\cite{Adams2013} (see the Supplemental Material, SM).


In this Letter, we compute the GW signal of a PBH-triggered SN~Ia from hydrodynamical simulations, and find that its linear memory is an order of magnitude larger than that of an SD explosion, which is instead ignited near the center~\cite{Falta2011, Seitenzahl:2015mdy, Seitenzahl:2012kr, Malone:2013tia}. A DD explosion, in contrast, is preceded by the GW signal from the inspiral of the binary~\cite{Korol2024, Korol:2021pun, Thiele:2021yyb, Sala:2025uqh, Das:2024quw}. We thus show that GWs, possibly corroborated by EM observations, can distinguish between the three progenitor channels, as summarized in Fig.~\ref{fig:smokinggun}.

\begin{figure*}
	\centering
	\includegraphics[width=0.4\linewidth]{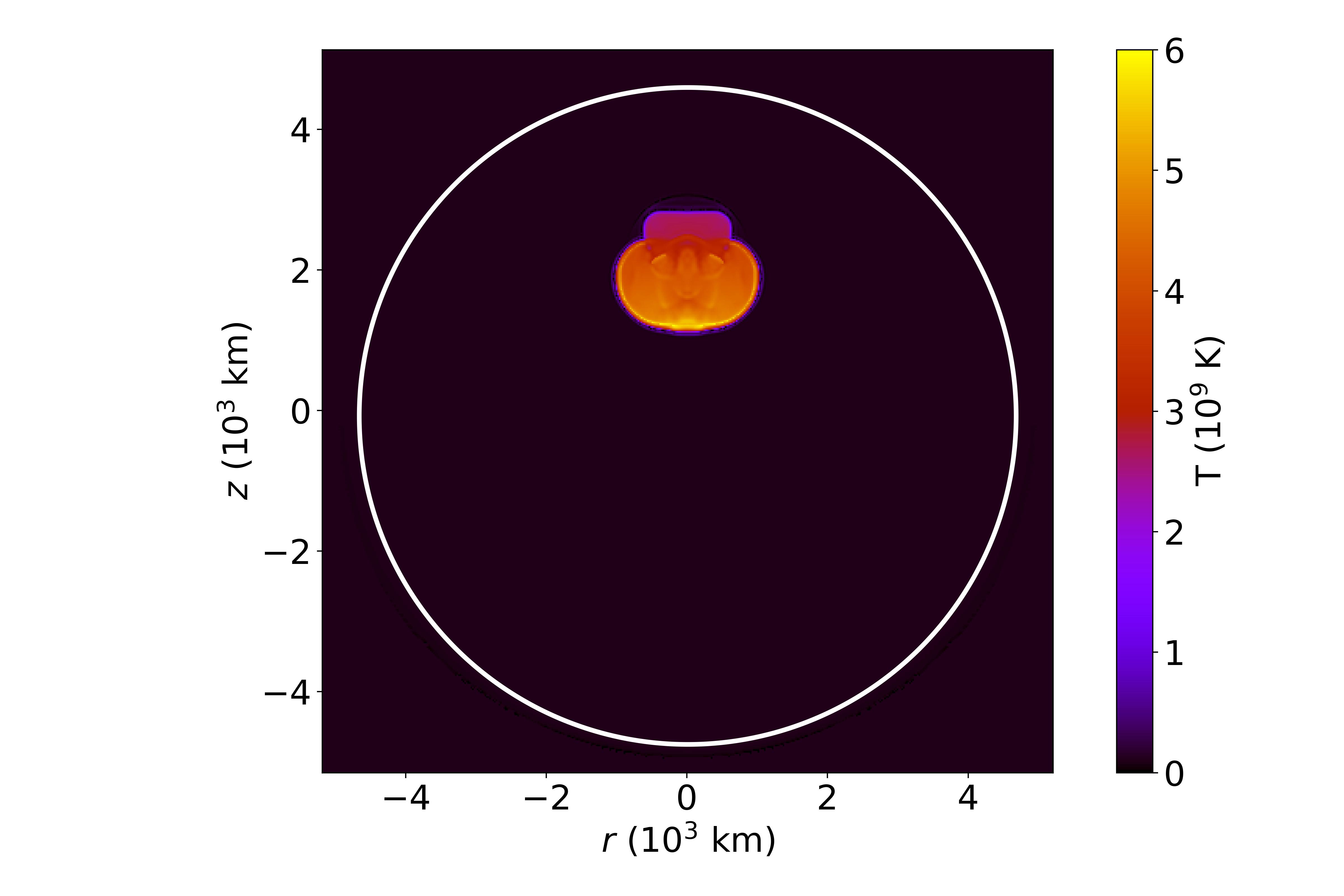}
	\includegraphics[width=0.4\linewidth]{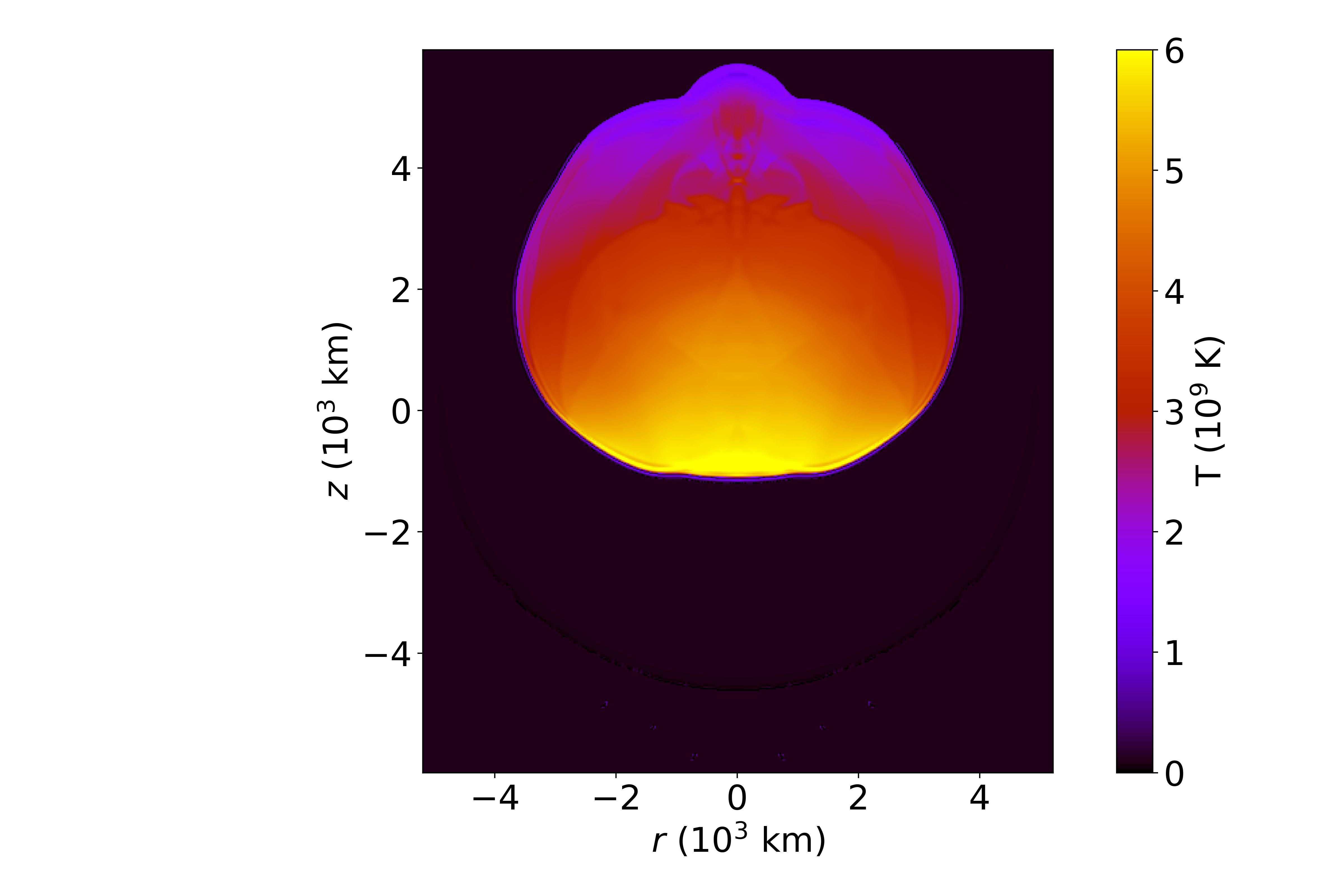}\\
	\includegraphics[width=0.4\linewidth]{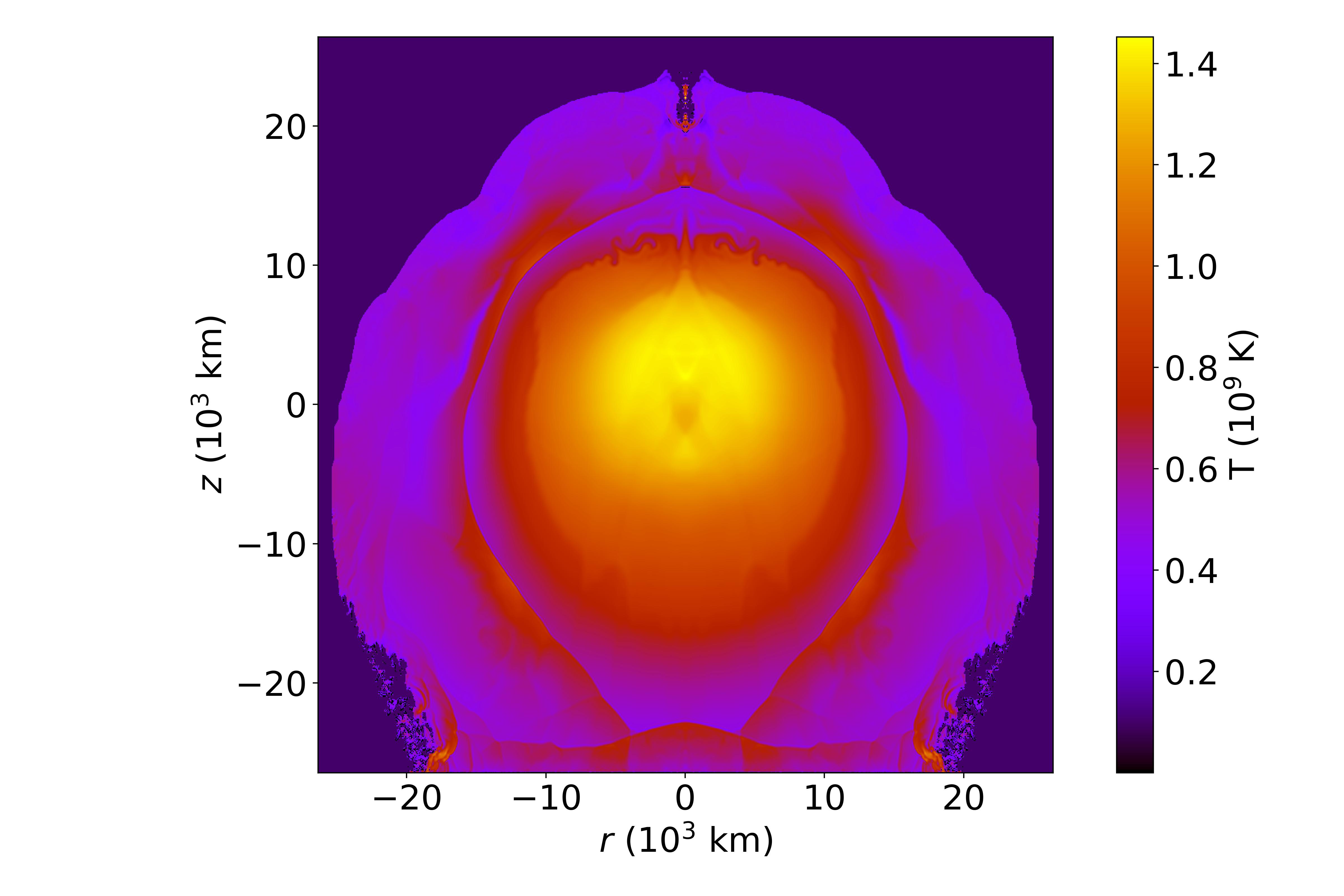}
	\includegraphics[width=0.4\linewidth]{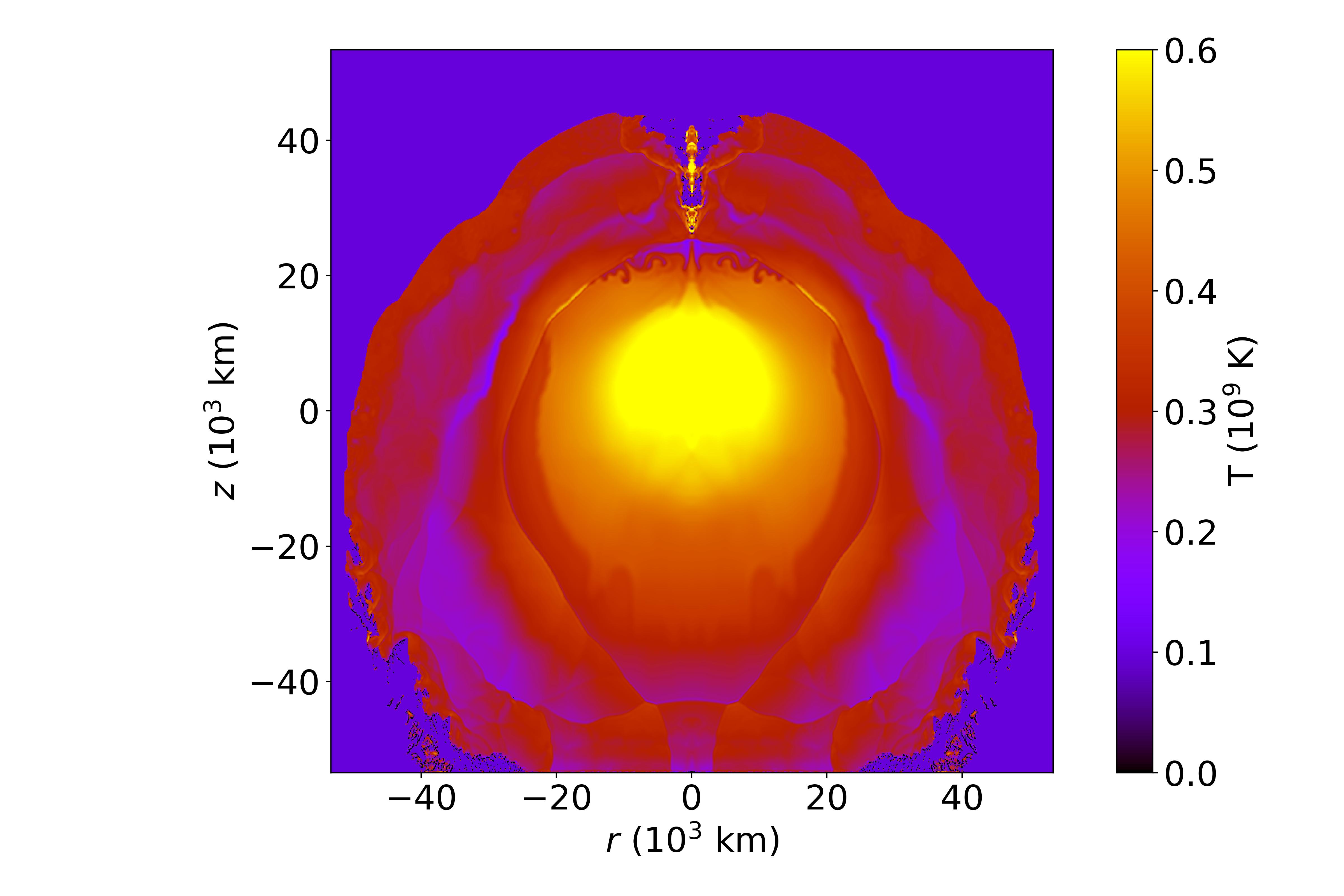}
	\caption{Temperature of the characteristic PBH-triggered explosion at 0.25 (top left), 0.50 (top right), 1.50 (bottom left) and 2.50~s (bottom right). The white circle marks the initial stellar radius. The asymmetry seeded by the off-center ignition survives into the elongated, homologously expanding ejecta. All panels refer to Model 06B-noKH of Ref.~\cite{Leung:2025enj}: $M_{\text{\tiny WD}}=1.08\,M_\odot$, $R_{\text{\tiny WD}}\simeq 5200\,{\rm km}$, $R_{\rm ign}\simeq 0.44\,R_{\text{\tiny WD}}$.}
	\label{fig:temp_profile}
\end{figure*}

\vspace{0.1cm}
\noindent{{\bf{\em Numerical simulations.}}}
The Type Ia supernova simulation is performed by solving the Euler equations in cylindrical coordinates \cite{Leung2015WENO}, coupled with sub-grid turbulence \cite{Clement1993SubGridTurb, Niemeyer1995TurbFlame}. The code uses the level-set method to track the deflagration and detonation fronts \cite{Osher1988, Reinecke1999LevelSet}, with the laminar flame with Rayleigh-Taylor acceleration prescription from \cite{Timmes1992, Hicks2015RTFLame}. The hydrodynamical simulations assume the \texttt{Helmholtz} Equation of State \cite{Timmes1999Helmholtz, Timmes2000Helmholtz}, which consists of electron gas at arbitrary relativistic and degenerate level, nuclei in the form of classical ideal gas, photon gas with blackbody distribution and $e^--e^+$ creation-annihilation pair correction. For the chemical composition, the code uses the 7-isotope nuclear reaction network \cite{Timmes2000Helmholtz, Timmes20007Iso, Timmes1999Torch} and the 3-step nuclear burning to track the detailed energy releases \cite{Calder2007NucTime}. We refer the interested reader to \cite{Leung2015WENO} for detailed code documentation and code test. The modular structure of the code allows us to adapt the code with various explosion physics, including the Chandrasekhar mass WD \cite{Leung2018Chand, Leung2020ChandIax} and sub-Chandrasekhar mass WD \cite{Leung2020ChandIax, Leung2020SubChand, Leung2021SN2014J}.

\begin{figure*}
	\centering
	\includegraphics[height=0.235\linewidth]{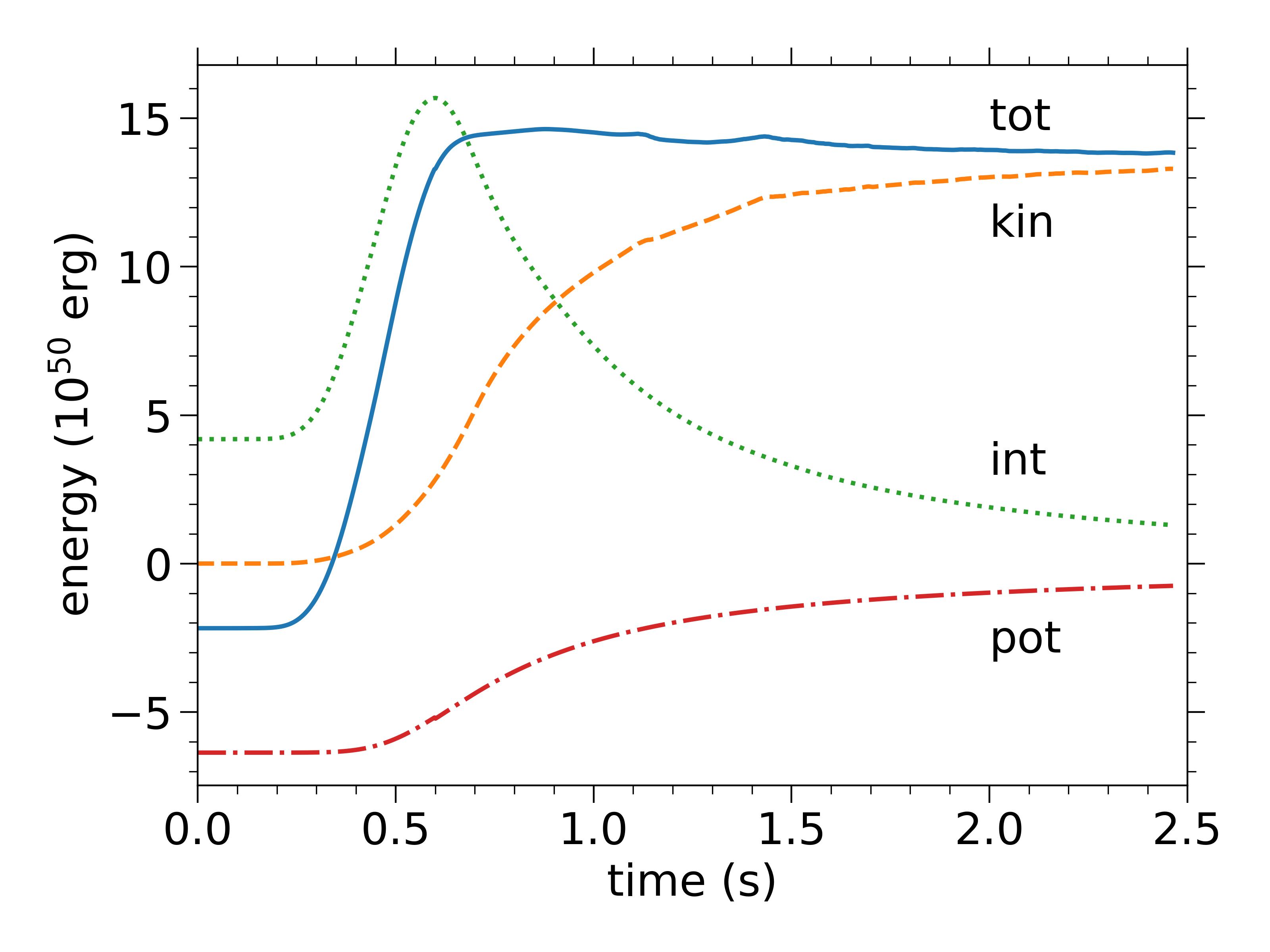}\hfill
	\includegraphics[height=0.235\linewidth]{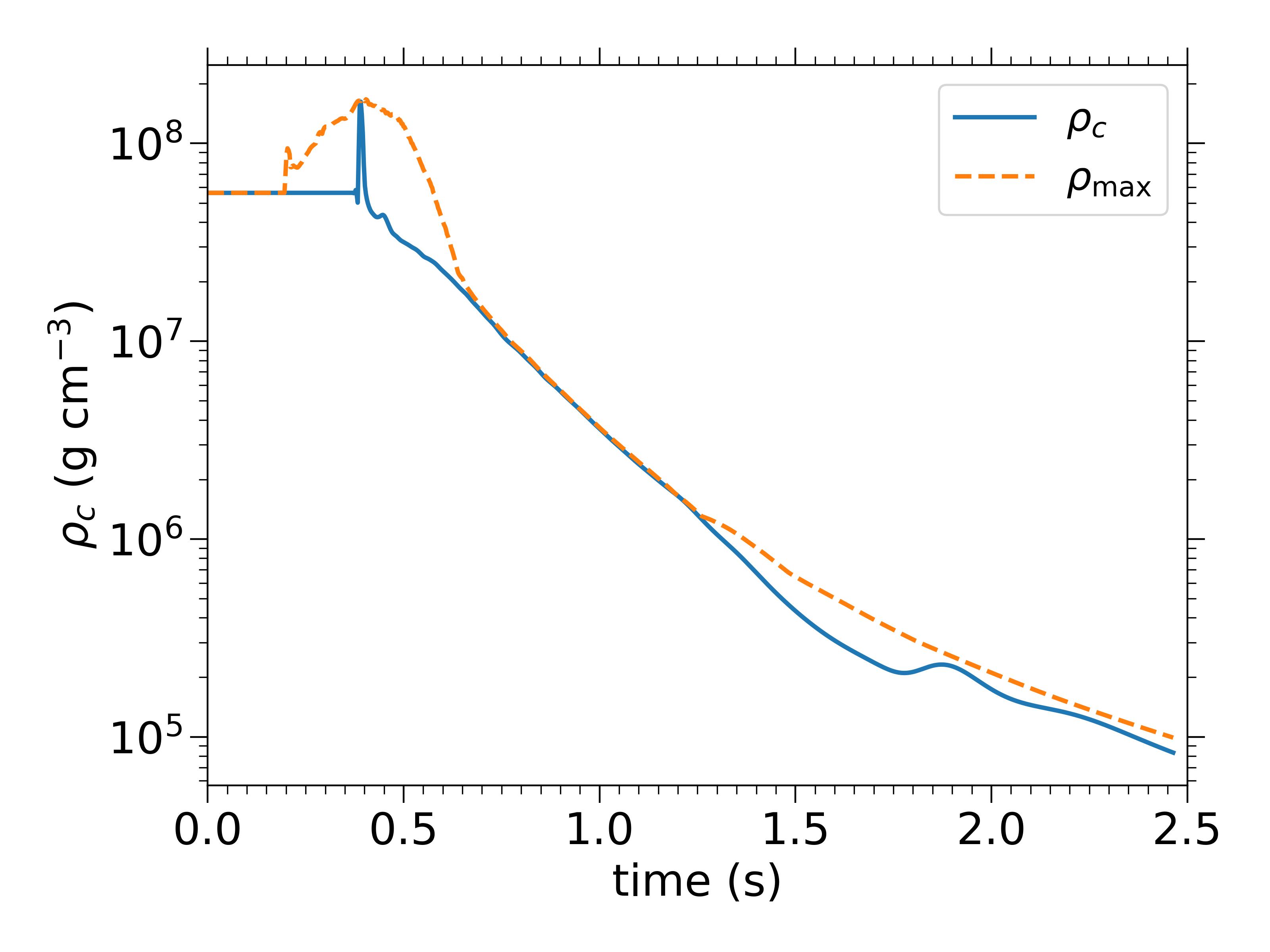}\hfill
	\includegraphics[height=0.235\linewidth]{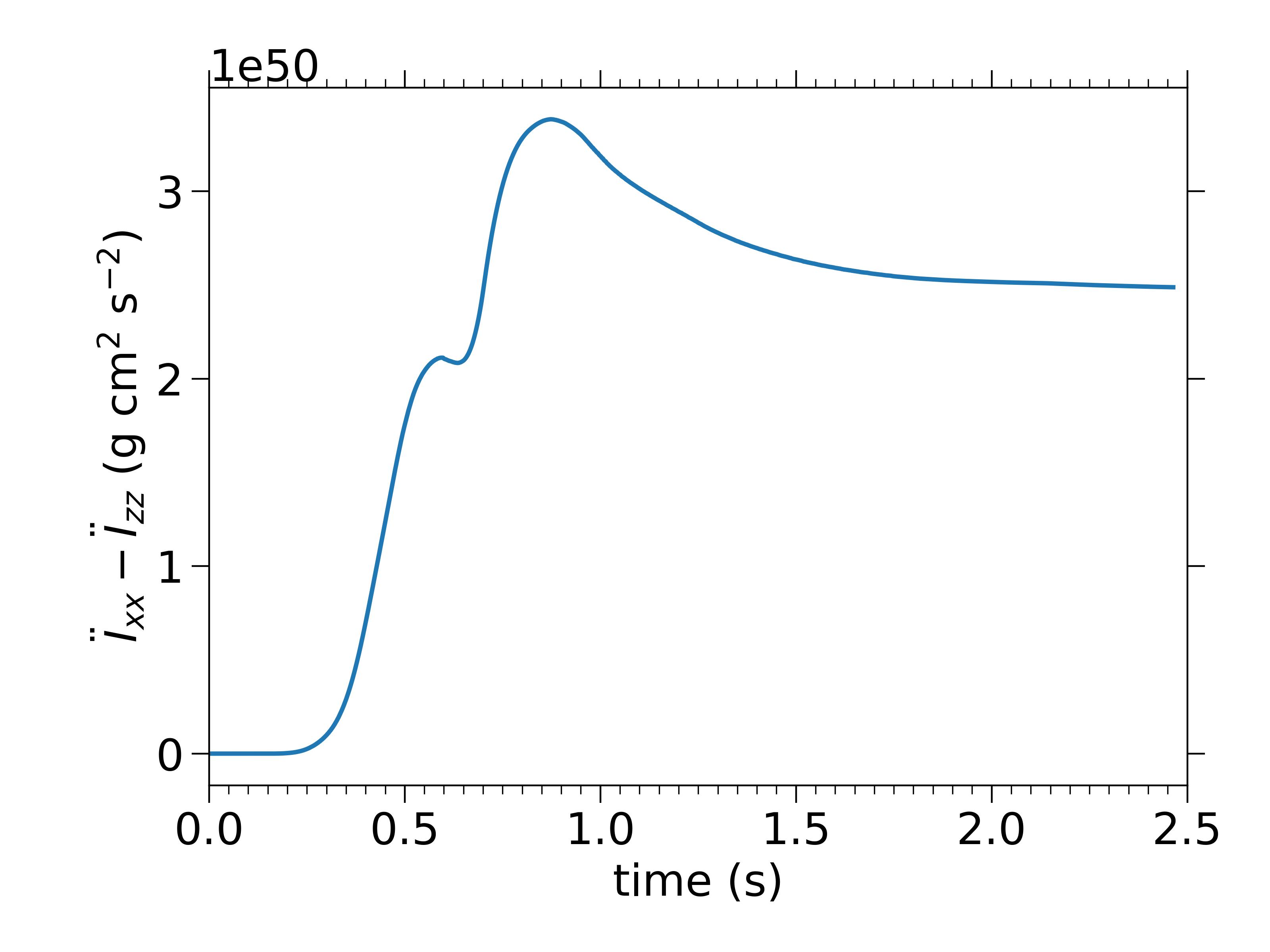}
	\caption{\textbf{\textit{Left panel:}} Total (solid), kinetic (dashed), internal (dotted) and potential (dot-dashed) energy of the simulated model. \textbf{\textit{Middle panel:}} Central density $\rho_c$ (solid) and global maximum density $\rho_{\rm max}$ (dashed). \textbf{\textit{Right panel:}} $\ddot I_{xx}-\ddot I_{zz}$ of the simulated model, from Eq.~\eqref{eq:Iddot}.}
	\label{fig:energy_rhoc}
\end{figure*}

For the explosion mechanism, we use the deflagration-detonation transition (DDT) with turbulent deflagration \cite{Roepke2007DDTInit, Roepke2007DDT}. The DDT criterion is evaluated along the deflagration front; where it is met, a detonation front is seeded and evolved independently~\cite{Sharpe1999}. The simulations follow the explosion and expansion of the WD until the matter reaches homologous expansion. We terminate the simulations once most chemical reactions cease when the ejecta become cold and low-density. 
For the progenitor, we use the ignition mass coordinate computed in \cite{MC19}. In that article, for a given progenitor C+O white dwarf, the enclosed mass coordinate at which the thermonuclear runaway can be spontaneously triggered is given, depending on whether Kelvin--Helmholtz instabilities efficiently cool the region heated by the PBH. In \cite{Leung:2025enj} we developed a sequence of models assuming a quadrant of the WD. In this work, we expand the calculation by computing the hemisphere of the WD (i.e., with rotational symmetry but no reflection symmetry). 

For the simulation in this paper, we consider a $1.08\,M_\odot$ C+O WD (model 06B-noKH of Ref.~\cite{Leung:2025enj}), with central density $\rho_c\simeq 5.6\times10^{7}\,{\rm g\,cm^{-3}}$ and radius $R_{\text{\tiny WD}}\simeq 5200\,{\rm km}$. 
We consider a single-bubble configuration in which the thermonuclear runaway is ignited on the rotational symmetry axis ($z$-axis) at the mass coordinate $M_{\rm ign}=0.6\,M_\odot$, i.e.\ $R_{\rm ign}\simeq 2280\,{\rm km}\simeq 0.44\,R_{\text{\tiny WD}}$. This mimics a PBH crossing the WD along a chord and igniting the runaway at the point of highest density along its trajectory. This is the realistic case: the energy lost to dynamical friction in a single transit, relative to the PBH energy at infinity, is~\cite{Gottlieb:2026kpg}
\begin{equation}
    \frac{\Delta E_{\rm df}}{E^\infty_\PBH}
    \sim
    C_{\mathcal{I}}
    \left(\frac{M_{\rm PBH}}{M_\WD}\right)
    \left(\frac{v_{\rm esc}}{v^\infty_\PBH}\right)^{2}\,,
\end{equation}
where $v^\infty_\PBH \sim 10^{-3}\,c$ is the asymptotic PBH velocity in the DM halo. Since $M_{\rm PBH}\ll M_\WD$, this ratio is tiny ($\sim 10^{-10}$ for the benchmark parameters of Eq.~\eqref{eq:cylinder}), so the PBH is not captured and crosses the star once~\cite{DeLorenci:2024xez}.

In Figure \ref{fig:temp_profile} we plot the snapshot of the temperature profiles of the explosion model considered at four timesteps, highlighting the different phases of the explosion including the homologous expansion phase, where the ejecta has a size about ten times the initial radius. The early asymmetry due to the explosion kernel persists throughout the simulation and is preserved as an elongated shape.

To characterize the overall explosion progress, in Figure \ref{fig:energy_rhoc} (left panel) we plot the total, kinetic, internal and potential energy of the explosion model. The exothermic processes during explosion take place within the first 0.5 s, where the matter is mostly burnt by supersonic detonation. It reaches a final energy of $1.5 \times 10^{51}$ erg, which is stored as internal energy. After that, the disrupted WD begins its expansion. The pressure work done converts the internal energy into kinetic energy, which is mostly converted after 1.5 s. At that point, the star has significantly expanded, where the potential energy approaches zero. In Fig.~\ref{fig:energy_rhoc} (middle panel), we plot the central (core) density and the global maximum density inside the WD. For a spherical WD, the core density is the highest across the star, which is true before the detonation wave arrives at the center around $t = 0.4$ s. When the shock wave propagates against the density gradient, it creates a density contrast large enough to supersede the core density, until it sweeps across the other side of the WD. This creates a density asymmetry and, hence, a time-varying quadrupole moment. 
The latter point is quantified in the right panel of Fig.~\ref{fig:energy_rhoc}, where we plot the difference between the second time derivatives of the diagonal components of the moment of inertia,
\begin{equation}\label{eq:Iddot}
	\ddot{I}_{xx} - \ddot{I}_{zz} = 2 \int_V \rho \left(v_z^2 - v_x^2 + x \frac{\partial\Phi}{\partial x} - z \frac{\partial\Phi}{\partial z} \right) dV.
\end{equation}
Note that the formula utilizes the Euler equation to rewrite the time derivative of the density in terms of hydrodynamics variable (see SM for the derivation). This avoids the finite differencing with respect to time which introduces additional noise in the high frequency domain~\cite{Finn1990, Blanchet1990}. The ejecta appear to be spherical at the beginning, and when the detonation converges, the onset of aspherical motion creates strong differences at $t = 0.5$ s and $0.8$ s.

\begin{figure*}[t]
	\centering
	\includegraphics[width=0.5\linewidth]{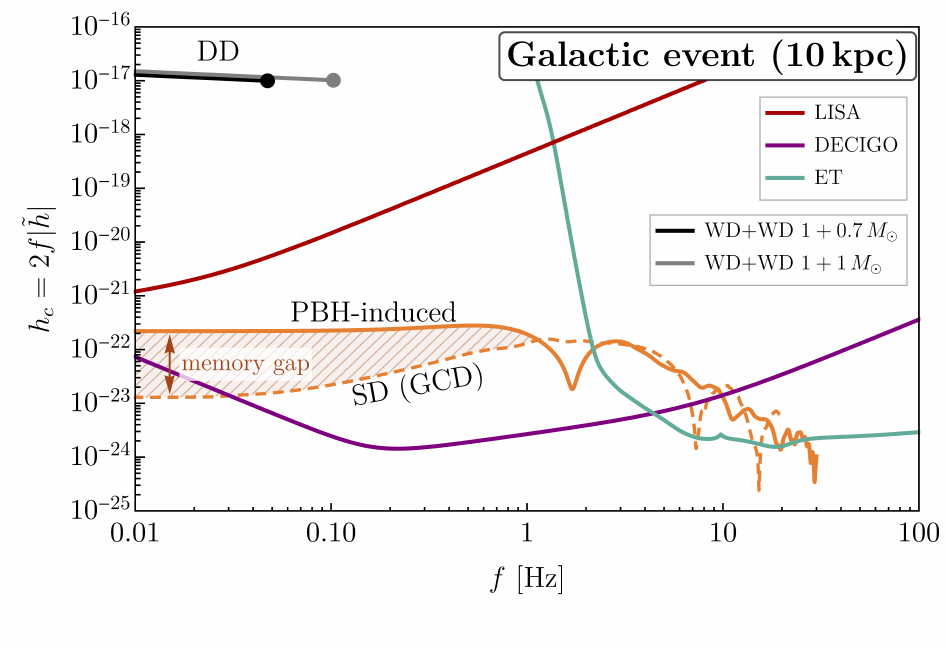}%
	\hfill
	\includegraphics[width=0.5\linewidth]{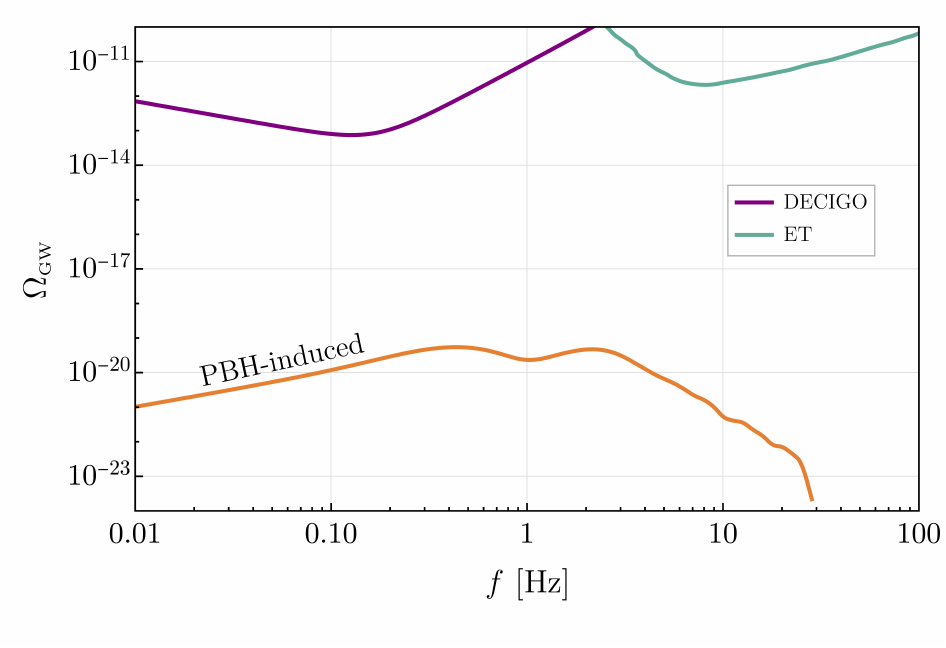}
	\caption{\textbf{\textit{Left panel:}} Characteristic strain of a Galactic event at $10\,{\rm kpc}$, viewed edge-on ($\iota=90^\circ$). Orange solid: PBH-triggered SN~Ia (this work). Orange dashed: SD gravitationally confined detonation of Ref.~\cite{Falta2011}. Both are continued below the simulated band by their memory plateau, $h_c\to\Delta h/\pi$; the hatched region marks the memory gap. Black and gray: inspiral of DD progenitors with $1+0.7\,M_\odot$ and $1+1\,M_\odot$, up to Roche-lobe overflow (dots). Also shown are the noise curves $h_n=\sqrt{f S_n}$ of LISA~\cite{Robson2019}, DECIGO and ET. \textbf{\textit{Right panel:}} Stochastic background $\Omega_{\rm GW}$ from the cosmological population, assuming all SNe~Ia are PBH-triggered, compared with the DECIGO and ET sensitivities.}
	\label{fig:money}
\end{figure*}

\vspace{0.1cm}
\noindent{{\bf{\em GWs from a PBH-triggered explosion}}}
In the quadrupole approximation, with $\iota$ the angle between the line of sight and the symmetry axis and $d_{\text{\tiny L}}$ the luminosity distance~\cite{Maggiore2008}, the emitted gravitational wave reads
\begin{equation}\label{eq:strain}
  h_+ = \frac{G}{c^4 d_{\text{\tiny L}}}\,\sin^2\iota \,\left(\ddot{I}_{xx}-\ddot{I}_{zz}\right), \qquad h_\times = 0\,.
\end{equation}
As a direct consequence of the imposed axisymmetry, the cross polarization vanishes and the emitted gravitational-wave signal is therefore linearly polarized. Once the ejecta enter the coasting phase ($t\gtrsim 1.5\,{\rm s}$), the kinetic energy dominates the energy budget (see the left panel of Fig.~\ref{fig:energy_rhoc}), the gravitational potential becomes negligible ($\Phi\to 0$), and the velocities freeze. Equation~\eqref{eq:Iddot} then tends to the constant value $\Delta\ddot I = 4 A\,E_{\rm kin}$, where
\begin{equation}\label{eq:kinetic_anisotropy}
	A\equiv\int\rho\,(v_z^2-v_x^2)\,{\rm d}^3x\big/\!\int\rho\, v^2\,{\rm d}^3x\,,
\end{equation}
is the kinetic anisotropy of the ejecta. A source that ejects matter anisotropically leaves a permanent offset $\Delta h$ in the gravitational-wave strain, known as linear memory~\cite{Zeldovich1974, Braginsky:1985vlg, Braginsky1987, Favata2010, Boybeyi:2024aax, Brabant2026}. It is formally defined as the net change in the metric perturbation between the final and initial states~\cite{Christodoulou1991, Thorne:1992sdb}, i.e.,
\begin{equation}\label{eq:memory_def}
	\Delta h \equiv\! \lim_{t\to+\infty} h_+(t) \! -\! \lim_{t\to-\infty} h_+(t) = \frac{4G}{c^4 d_{\text{\tiny L}}} A\, E_{\rm kin}\sin^2\iota\,.
\end{equation}
 In the Fourier domain, the memory signal accumulates at  low frequencies. This can be understood with a simple toy model: modeling the rise with a logistic function
\begin{equation}
	h(t)=\frac{\Delta h}{\left(1+e^{-t/\tau}\right)}\,,
\end{equation}
where $\tau$ sets the rise time and $\Delta h$ the amplitude of the memory, one finds the characteristic strain
\begin{equation}\label{eq:hc}
	h_c(f)\equiv 2f\,|\tilde h(f)| = \frac{\Delta h}{\pi}\,\frac{x}{\sinh x}\,,\qquad x\equiv 2\pi^2 f\tau\,.
\end{equation}
For $f\ll 1/\tau$, the characteristic strain is flat, $h_c\to\Delta h/\pi$, while for $f\gg 1/\tau$ it is exponentially suppressed. A low-frequency plateau in the characteristic strain is therefore the distinctive signature of the memory. This behavior is evident in the left panel of Fig.~\ref{fig:money}, where the solid orange line shows the characteristic strain computed from our numerical simulation of a PBH-induced WD explosion assuming $\iota = 90^\circ$ (edge-on). The spectrum exhibits a knee at $f\sim 1\,{\rm Hz}$, corresponding to a rise time $\tau\sim 1\,{\rm s}$, consistent with the energy and strain evolution shown in Fig.~\ref{fig:energy_rhoc}. For $f \ll 1\, {\rm Hz}$ we read $h_c \to 2 \times 10^{-22}$, corresponding to a memory amplitude $\Delta h \simeq \pi \,h_c(f \ll 1\, {\rm Hz}) \simeq 7 \times 10^{-22}$ at $d_{\text{\tiny L}} = 10 \,{\rm kpc}$, or $d_{\text{\tiny L}}\,\Delta h \simeq 22 \,{\rm cm} $. Given that the asymptotic kinetic energy is $E_{\rm kin}\simeq 1.3\times10^{51}\,{\rm erg}$ (see the left panel of Fig.~\ref{fig:energy_rhoc}), we use Eq.~\eqref{eq:memory_def} to infer the kinetic anisotropy of the ejecta from our simulation, obtaining $A\simeq 0.05$. We have also computed the kinetic anisotropy for two other simulations listed in Ref.~\cite{Leung:2025enj}, namely 02B-noKH and 10B-noKH; the results are shown in Tab.~\ref{tab:rign_A}. We find that increasing the ratio $R_{\text{\tiny{ign}}}/R_\WD$ leads to an even larger asymmetry; we discuss the physical origin of this trend below.

\begin{table}
  \centering
  \begin{tabular}{lccc}
    \hline
    Model & 02B-noKH & 06B-noKH & 10B-noKH \\
    \hline
    $A$ & 0.01 & 0.05 & 0.1 \\
    $R_{\text{\tiny{ign}}}/R_\WD$ & 0.285 & 0.438 & 0.542 \\
    \hline
  \end{tabular}
    \caption{Kinetic anisotropy of the ejecta $A$ as a function of the ignition radius relative to the stellar radius, $R_{\text{\tiny{ign}}}/R_\WD$.}
    \label{tab:rign_A}
\end{table}

\vspace{0.1cm}
\noindent{{\bf{\em Progenitor identification.}}}
In the left panel of Fig.~\ref{fig:money}, we also show the characteristic strain from the numerical simulation of Ref.~\cite{Falta2011}, in which the authors investigated the single-degenerate (SD) channel. In this case, a near-Chandrasekhar-mass WD ignites close to its center: their three-dimensional gravitationally confined detonation (GCD) model starts from a bubble $40\,{\rm km}$ off-center in a $1.37\,M_\odot$ WD ($R_{\text{\tiny ign}}/R_{\text{\tiny WD}}\sim 0.02$). The bubble rises and breaks out, and the detonation, triggered at the antipode, sweeps over the still-bound, nearly spherical star. This produces a sharp burst with a peak strain $d_{\text{\tiny{L}}}\,h_{\rm peak}\simeq 12\,{\rm cm}$, comparable to the PBH-induced case. The two scenarios also have comparable final kinetic energies ($E_{\rm kin}\simeq 1.4\times10^{51}\,{\rm erg}$ in the GCD simulation), yet the memory is much smaller, $d_{\text{\tiny{L}}}\,\Delta h\simeq 1.2\,{\rm cm}$, corresponding to a kinetic anisotropy $A\simeq 3\times10^{-3}$. In Ref.~\cite{Seitenzahl:2015mdy}, the authors compute the GW signal (not shown in Fig.~\ref{fig:money} to avoid clutter) from a further three-dimensional hydrodynamical simulation, again in the SD channel, but using a deflagration-to-detonation transition (DDT) model with ignition within $\sim 100\,{\rm km}$ of the center. Although the final kinetic energy is again comparable to that of the PBH-triggered explosion, the linear memory is comparable to that of the GCD model of Ref.~\cite{Falta2011}, corresponding again to a kinetic anisotropy $A\sim 10^{-3}$. We therefore argue that a large memory is a fossil of the ignition geometry, rather than a feature of the particular explosion mechanism (GCD or DDT). In the SD channel, ignition is constrained to occur close to the center, since it results from stochastic fluctuations of the fuel density there. A PBH, instead, ignites the star far off-center, e.g., at $R_{\text{\tiny ign}}\simeq 0.44\,R_{\text{\tiny WD}}$ in our simulation. By the time of the DDT, the star is already distorted, and the detonation freezes this asymmetry into elongated ejecta (Fig.~\ref{fig:temp_profile}). As a result, the memory is one order of magnitude larger than in the SD scenarios, as highlighted by the memory gap in Fig.~\ref{fig:money} (left panel). Since both signals scale identically with $\iota$ and $d_{\text{\tiny L}}$, this ratio is independent of the orientation and distance of the source. In the double-degenerate (DD) channel, the GW signal from the WD explosion is preceded by the GWs emitted during the inspiral of the WD binary~\cite{Ebadi:2024oaq}. The corresponding characteristic strain is shown in the left panel of Fig.~\ref{fig:money} for a $1+0.7\,M_\odot$ binary (black) and a $1+1\,M_\odot$ binary (gray). The dots mark the point along the inspiral one day before the binary reaches the Roche frequency, after which the explosion is expected to occur, as found in the simulation of Ref.~\cite{2022MNRAS.517.5260P}. The inspiral signal is in the frequency range $10^{-2}-10^{-1} \, {\rm Hz}$ and is orders of magnitude larger than the explosion strain. It is also detectable by LISA for a Galactic event. We therefore conclude that, should a Galactic Type~Ia supernova occur in the near future, GWs alone could identify the likely progenitor channel. Complementary electromagnetic (EM) observations, in particular of signatures associated with the non-degenerate companion in the SD channel, would provide further information to corroborate this conclusion~\cite{Howell:2015xzw}.

\vspace{0.1cm}
\noindent{{\bf{\em Detection prospects and stochastic background.}}}
The detectability of a single event is quantified by the optimal (matched-filter) signal-to-noise ratio (SNR)~\cite{Finn:1992xs, Sutton:2013ooa}. For a detector with one-sided noise power spectral density $S_n(f)$, it reads
\begin{equation}\label{eq:snr}
	\rho^2 = 4 \int_0^\infty \frac{|\tilde{h}(f)|^2}{S_{n}(f)} \, \d f = \int \left[\frac{h_c(f)}{h_n(f)}\right]^2 \d\ln f\,,
\end{equation}
where in the second equality we have used the characteristic strain of Eq.~\eqref{eq:hc} and the noise amplitude $h_n\equiv\sqrt{f S_n}$~\cite{Moore:2014lga}, i.e., the two quantities shown in the left panel of Fig.~\ref{fig:money}. 

Using the analytic DECIGO noise curve of Ref.~\cite{Yagi2011}, the ET2026a baseline~\cite{ET_SensitivityCurves_2026a, ET_DesignTaskForce:2025, Dupletsa:2026ET} and Cosmic Explorer (CE)~\cite{Evans2021} curves, and the LIGO O4 design curve~\cite{KAGRA:2021tnv, KAGRA:2025yfg, LIGOScientific:2026bxu}, we find for the PBH-triggered event at $10\,{\rm kpc}$, viewed edge-on, $\rho\simeq 230$ for DECIGO, $12.0$ for the Einstein Telescope (ET), $3.0$ for CE, and $0.02$ for LIGO O4. DECIGO, whose sensitivity peaks at $0.1$--$1\,{\rm Hz}$, collects its SNR from the memory plateau. Ground-based detectors, whose sensitivity degrades steeply below a few hertz, only see the exponentially suppressed high-frequency tail of the rise ($x\gg 1$ in Eq.~\eqref{eq:hc}). Since $\rho\propto d_{\text{\tiny L}}^{-1}$, an event is detected above a threshold $\rho_{\rm th}$ out to $d_{\rm max}=10\,{\rm kpc}\times\rho(10\,{\rm kpc})/\rho_{\rm th}$. For $\rho_{\rm th}=8$, DECIGO reaches $d_{\rm max}\simeq 280\,{\rm kpc}$, which covers the whole Milky Way, the Magellanic Clouds and most of the Galactic satellites, whereas ET requires an event within $\simeq 15\,{\rm kpc}$---well inside the Milky Way. Complementary electromagnetic observations of the Type~Ia supernova, together with a comparison between the observed peak GW strain and its numerically simulated value ($d_{\text{\tiny{L}}}\,h_{\rm peak}/\sin^2 \iota \simeq 12\,{\rm cm}$), help break the degeneracy between the luminosity distance and the unknown inclination angle $\iota$.
%
%
A Galactic SN~Ia is, however, a rare gift. Treating explosions as a Poisson process with the Galactic rate $\mathcal{R}_{\rm MW}\simeq(3$--$7)\times10^{-3}\,{\rm yr^{-1}}$~\cite{Korol2024, Adams2013}, the probability of at least one event during an observing time $T$ is $1-e^{-\mathcal{R}_{\rm MW}T}$. This gives $1$--$3\%$ for a $T\simeq 4\,{\rm yr}$ DECIGO mission, $3$--$7\%$ for LISA ($T\simeq 10\,{\rm yr}$), which would observe the inspiral of a DD progenitor, and $14$--$30\%$ over the $\sim 50\,{\rm yr}$ lifetime of ET. 

More distant events add up incoherently to a stochastic GW background (SGWB)~\cite{Lei:2025jsu}. To compute it, we first need the energy spectrum of a single explosion. For an axisymmetric source $h_+\propto\sin^2\iota$ in Eq.~\eqref{eq:strain}, and integrating the quadrupole flux $\d E_{\rm GW}/(\d A\,\d f)=(\pi c^3/2G)\,f^2|\tilde h|^2$ over the sphere gives
\begin{equation}\label{eq:dEdf}
	\frac{\d E_{\rm GW}}{\d f}=\frac{16\pi^2}{15}\,\frac{c^3}{G}\left[d_{\text{\tiny L}}\,f\,|\tilde h_0(f)|\right]^2\;\xrightarrow{f\to0}\;\frac{4G}{15c^5}\left(\Delta\ddot I\right)^2\,,
\end{equation}
where $\tilde h_0$ is the edge-on spectrum. The low-frequency limit follows from the memory plateau, $f|\tilde h_0|\to\Delta h/2\pi$, together with $d_{\text{\tiny L}}\Delta h=(G/c^4)\,\Delta\ddot I$ for an edge-on observer, and evaluates to $\simeq 5\times10^{40}\,{\rm erg\,Hz^{-1}}$ for our model. Integrating Eq.~\eqref{eq:dEdf} over frequency gives the radiated energy $E_{\rm GW}\simeq 1.1\times10^{41}\,{\rm erg}$, in agreement to within $0.2\%$ with the time-domain value $(2c^3/15G)\int(d_{\text{\tiny L}}\,\dot h_0)^2\,\d t$. This is a fraction $\sim 10^{-10}$ of the kinetic energy of the ejecta, a small efficiency that reflects the low compactness of the WD.  Assuming that all SNe~Ia are PBH-triggered (see SM, Sec.~S2), the energy density of the SGWB, normalized to the critical density $\rho_c=3H_0^2/(8\pi G)$, is then
\begin{equation}\label{eq:OmegaGW}
	\Omega_{\rm GW}(f)=\frac{f}{\rho_c \,c^2}\int_0^{z_{\rm max}}\!\d z\,\frac{R_{\rm Ia}(z)}{(1+z)H(z)}\,\frac{\d E_{\rm GW}}{\d f_s}\bigg|_{f_s}\,,
\end{equation}
where $f_s=(1+z)f$ is the source-frame frequency and the factor $1/(1+z)$ accounts for cosmological time dilation. For the comoving SN~Ia rate density we take $R_{\rm Ia}=2.5\times10^{-5}(1+z)^{1.5}\,{\rm Mpc^{-3}\,yr^{-1}}$ for $z<1$ and $9.7\times10^{-5}(1+z)^{-0.5}\,{\rm Mpc^{-3}\,yr^{-1}}$ for $1\leq z\leq z_{\rm max}=2$~\cite{Dilday2008, Hounsell2018, Mitra2025}. Since $\d E_{\rm GW}/\d f_s$ is flat on the memory plateau, $\Omega_{\rm GW}$ grows linearly with $f$ below the redshifted knee, giving
\begin{equation}\label{eq:Omega_est}
	\Omega_{\rm GW}(f)\simeq 10^{-19}\left(\frac{d_{\text{\tiny L}}\Delta h}{22\,{\rm cm}}\right)^2\left(\frac{f}{1\,{\rm Hz}}\right)\,,\qquad f\lesssim 1\,{\rm Hz}\,,
\end{equation}
while the spectrum is exponentially suppressed at higher frequencies (see Fig.~\ref{fig:money}, right panel). A SGWB is searched for by cross-correlating the outputs of two detectors, with $\rho_{\rm bg}^2=2T\int\d f\,[\Omega_{\rm GW}(f)/\Omega_{\rm sens}(f)]^2$. For DECIGO, we find $\rho_{\rm bg}\simeq 9\times10^{-4}$ in $T=1\,{\rm yr}$ and $\simeq 2\times10^{-3}$ in $T=4\,{\rm yr}$, three orders of magnitude below detection. Physically, the larger number of cosmological sources does not compensate for the $d_{\text{\tiny L}}^{-2}$ dilution of their flux. The memory of PBH-triggered SNe~Ia is therefore a target for individual Galactic and Local Group events, rather than for SGWB searches.

\vspace{0.1cm}
\noindent{{\bf{\em Conclusions.}}}
GW observations of a Galactic SN~Ia can identify its progenitor. A DD system is announced by its milli- to deci-Hertz inspiral, which precedes the explosion and is orders of magnitude louder than the resulting GW burst. SD and PBH-triggered explosions, which have no GW precursor, differ by the presence of a non-degenerate companion and by the amplitude of the GW memory. Since $d_{\text{\tiny L}}\Delta h=(4G/c^4)\,A\,E_{\rm kin}\sin^2\iota$ and the final kinetic energy, $E_{\rm kin}\simeq(1.3$--$1.4)\times10^{51}\,{\rm erg}$, is fixed by nuclear burning in both channels, the memory directly measures the kinetic anisotropy $A$ of the ejecta, which in turn records the ignition geometry. From two-dimensional axisymmetric hydrodynamical simulations, we find that the off-center ignition by a PBH at $R_{\text{\tiny ign}}\simeq 0.44\,R_{\text{\tiny WD}}$ distorts the star before the detonation, which freezes this asymmetry into elongated ejecta with $A\simeq 0.05$. The resulting memory, $d_{\text{\tiny L}}\Delta h\simeq 22\,{\rm cm}$, is a factor $\simeq 18$ above that of centrally ignited SD explosions, $A\sim 10^{-3}$, irrespective of whether the latter proceed through a GCD or a DDT. The memory produces a flat characteristic strain below $\sim 1\,{\rm Hz}$, which DECIGO detects with $\rho\simeq 230$ at $10\,{\rm kpc}$ and out to $\simeq 280\,{\rm kpc}$, whereas ET and CE are limited to events within $\lesssim 15\,{\rm kpc}$. The contribution of the cosmological population to the SGWB, $\Omega_{\rm GW}\lesssim 10^{-19}$, is instead far below the reach of any planned detector~\cite{Yagi2011, LGWA} (see also Ref.~\cite{Falta:2011aa}).

Our results rely on axisymmetric models, which capture only the $m=0$ component of the quadrupole. Extending our analysis to three-dimensional simulations would capture the contribution of the non-axisymmetric ($m\neq0$) components, which could in principle lead to even larger asymmetries. Furthermore, computing the memory over a grid of ignition radii (corresponding to different PBH impact parameters) and WD masses would extend the results in Tab.~\ref{tab:rign_A}, allowing us to map it across the progenitor parameter space and to build a template bank for future searches. Finally, simulating the actual trajectory of the PBH through the WD, rather than assuming a single ignition site, would allow us to better constrain the ignition physics and the ranges of PBH and WD masses for which an explosion can occur. We leave these
considerations to future work.


\vspace{0.1cm}
\noindent{{\bf{\em Acknowledgments.}}}
We thank V.~De~Luca for useful discussions and for comments on the manuscript.
L.D.G. is supported by NSF Grants
No. AST-2606672, No. PHY-2513337, No. PHY-090003,
and No. PHY-20043, by the Simons Foundation [MPS-
SIP-00001698, E.B.], by the Simons Foundation International [SFI-MPS-BH-00012593-02]. A.K.\ was supported by the U.S.\ Department of Energy (DOE) under Grant No.\ DE-SC0009937 and by the World Premier International Research Center Initiative (WPI), MEXT, Japan (Kavli IPMU).
S.C.L. acknowledges the support of the NSF Grant No. AST-2316807. S.R. is supported in part by the U.S. National Science Foundation (NSF) under Grant No. PHY-1818899. S.R. is also supported by the Simons Investigator Grant
No. 827042.

\bibliography{refs}

\clearpage
\onecolumngrid
\setcounter{equation}{0}
\setcounter{figure}{0}
\setcounter{table}{0}
\renewcommand{\theequation}{S\arabic{equation}}
\renewcommand{\thefigure}{S\arabic{figure}}
\renewcommand{\thetable}{S\arabic{table}}

\begin{center}
    {\large\bf Supplemental Material}
\end{center}

\section{S1. Quadrupole from the hydrodynamical variables}
We derive Eq.~\eqref{eq:Iddot} following the approach of Refs.~\cite{Finn1990, Blanchet1990} (see also Eq.~(10.76) of Ref.~\cite{Maggiore2008}). The inertia tensor is $I_{ij}=\int_V\rho\,(r^2\delta_{ij}-x_ix_j)\,d^3x$, and axial symmetry about the $z$-axis enforces $I_{xx}=I_{yy}$. Explicitly,
\begin{equation}
I_{xx}=\int_V\rho\,(y^2+z^2)\,d^3x\,,\qquad I_{zz}=\int_V\rho\,(x^2+y^2)\,d^3x\,,\qquad I_{xx}-I_{zz}=\int_V\rho\,(z^2-x^2)\,d^3x\,.
\end{equation}
The strain along a line of sight at angle $\iota$ from the $z$ axis is, in the transverse-traceless gauge,
\begin{equation}
h^{\rm TT}_{ij}=\frac{G}{c^4 d_{\text{\tiny L}}}\,\sin^2\iota\,\left(\ddot I_{xx}-\ddot I_{zz}\right)\,{\rm diag}\{1,-1,0\}\,,
\end{equation}
which gives Eq.~\eqref{eq:strain}. We now use the continuity and Euler equations of Newtonian hydrodynamics,
\begin{equation}
\partial_t\rho+\partial_i(\rho v^i)=0\,,\qquad \rho\,\partial_t v^i+\rho\, v^k\partial_k v^i=-\partial^i p-\rho\,\partial^i\Phi\,.
\end{equation}
Taking a volume $V$ that encloses all the matter ($\rho=0$ on $\partial V$), the continuity equation and an integration by parts give
\begin{equation}
\dot I_{xx}-\dot I_{zz}=-\int_V \partial_i(\rho v^i)\,(z^2-x^2)\,d^3x=\int_V\rho\, v^i\,\partial_i(z^2-x^2)\,d^3x=2\int_V\rho\,(z\,v_z-x\,v_x)\,d^3x\,.
\end{equation}
Combining the two equations, the momentum density obeys
\begin{equation}
\partial_t(\rho v^i)=-\partial_k(\rho\, v^k v^i)-\partial^i p-\rho\,\partial^i\Phi\,.
\end{equation}
Therefore
\begin{align}
\ddot I_{xx}-\ddot I_{zz}&=2\int_V\left[z\,\partial_t(\rho v_z)-x\,\partial_t(\rho v_x)\right]d^3x\nonumber\\
&=2\int_V\Big[-z\,\partial_k(\rho v^k v_z)-z\,\partial_z p-z\rho\,\partial_z\Phi+x\,\partial_k(\rho v^k v_x)+x\,\partial_x p+x\rho\,\partial_x\Phi\Big]d^3x\,.
\end{align}
Integrating by parts, $\int z\,\partial_k(\rho v^kv_z)=-\int\rho v_z^2$ and $\int z\,\partial_z p=-\int p$, and similarly for $x$. The two pressure terms cancel, leaving
\begin{equation}\label{eq:Iddot_SM}
\ddot I_{xx}-\ddot I_{zz}=2\int_V\rho\left[\left(v_z^2-v_x^2\right)+\left(x\,\partial_x\Phi-z\,\partial_z\Phi\right)\right]d^3x\,,
\end{equation}


\section{S2. Rate of PBH-triggered explosions in the Milky Way}
We estimate the rate of ignition-triggering PBH--WD encounters in the Milky Way by integrating over the Galactic WD population, the PBH phase-space distribution and the encounter kinematics. The aim is to show that, if PBHs are the DM, they naturally produce SNe~Ia at the observed rate.

We model the WD number density as the sum of a disk and a bulge component~\cite{Napiwotzki:2009tf},
\begin{equation}\label{eq:nWD}
    n_{\WD}(R,z) = n_{\WD,\odot}\, e^{-\frac{R-R_\odot}{R_{\rm d}}}\, \text{sech}^2\!\left(\frac{z-z_\odot}{z_{\rm d}}\right) + \frac{C_{\rm b}}{r\,(r+r_{\rm b})^3}\,,
\end{equation}
where $r=\sqrt{R^2+z^2}$. The disk has an exponential radial profile and an isothermal vertical distribution, with scale radius $R_{\rm d} = 2.5 \, {\rm kpc}$, scale height $z_{\rm d} = 0.3 \, {\rm kpc}$ and solar position $(R_\odot,z_\odot)=(8.2,\,0.0208)\,{\rm kpc}$~\cite{Bland-Hawthorn:2016lwg}. The local WD density~\cite{Jimenez-Esteban2018, OBrien2024}, $n_{\WD,\odot} \simeq 4.49 \times 10^{-3} \,{\rm pc}^{-3}$, is taken from the $20\,{\rm pc}$ sample of the second Gaia data release~\cite{Hollands2018}. The bulge follows a Hernquist profile of scale radius $r_{\rm b}=1\,{\rm kpc}$, with normalization $C_{\rm b}=(f\,M_{\rm b}/2\pi)\times10^{-9}\simeq0.18\,{\rm pc^{-3}\,kpc^{3}}$. This is obtained by populating a bulge of mass $M_{\rm b}=10^{10}\,M_\odot$ with the WD-to-mass ratio $f=n_{\WD,\odot}/\rho_{\star,\odot}\simeq0.11\,M_\odot^{-1}$, where $\rho_{\star,\odot}=0.04\,M_\odot\,{\rm pc}^{-3}$ is the local disk mass density~\cite{Kroupa:2002ky}. The WD mass distribution $p_\WD(M_\WD)$ is a three-component Gaussian mixture peaked near $0.6\,M_\odot$~\cite{Kepler2015, Kepler:2006ns, Cunningham2024}, normalized to unity.

\begin{figure}[h]
    \centering
    \includegraphics[width=0.5\linewidth]{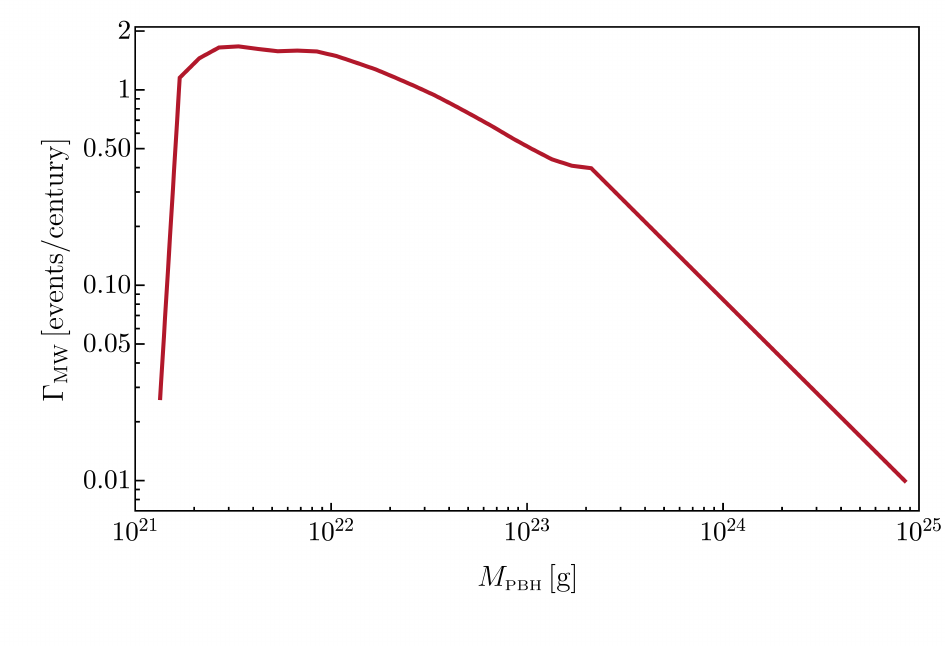}
    \caption{Rate of PBH-triggered WD explosions in the Milky Way (disk and bulge) as a function of the PBH mass, for $f_{\rm PBH}=1$.}
    \label{fig:rate}
\end{figure}

We assume that PBHs make up a fraction $f_{\rm PBH}$ of the DM with a monochromatic mass function of mass $m$, so that $n_{\rm PBH}(r)=f_{\rm PBH}\,\rho_{\rm DM}(r)/m$, with $\rho_{\rm DM}$ the Galactic halo profile assumed to follow a Navarro-Frenk-White (NFW) profile. Relative velocities follow a Maxwellian truncated at the escape speed,
\begin{equation}
  f(v)=\sqrt{\tfrac{2}{\pi}}\,\frac{v^2}{\sigma_v^3}\,
  e^{-v^2/2\sigma_v^2}\,,
\end{equation}
with dispersion $\sigma_v$ set by the halo. An encounter occurs when the PBH transits the WD, with geometric cross section enhanced by gravitational focusing~\cite{Mouri:2002mc},
\begin{equation}\label{eq:sigma}
  \sigma(M_{\rm WD},v)=\pi R_{\rm WD}^2
  \left[1+\frac{v_{\rm esc}^2}{v^2}\right],
\end{equation}
where $v_{\rm esc}^2=2GM_{\rm WD}/R_{\rm WD}$ and $R_{\rm WD}(M_{\rm WD})$ is the zero-temperature mass--radius relation. An explosion further requires the WD mass to lie in the ignitable range $M_{\rm WD}\in[0.75,1.38]\,M_\odot$ and the PBH mass to satisfy $m>m_{\rm min}(M_{\rm WD})$, where the minimum PBH mass for ignition $m_{\rm min}(M_{\rm WD})$ is digitized from Ref.~\cite{MC19} assuming no Kelvin--Helmholtz instabilities. Averaging Eq.~\eqref{eq:sigma} over $f(v)$ gives the rate per WD, $\langle\sigma v\rangle$, and the total Galactic rate is
\begin{equation}
  \Gamma_{\rm MW}=f_{\rm PBH}\!\int\! d^3r\, n_{\rm WD}(r)\,\frac{\rho_{\rm DM}(r)}{m}
  \int_{0.75\,M_\odot}^{1.38\,M_\odot}\! dM_{\rm WD}\,p_{\rm WD}(M_{\rm WD})\,\langle\sigma v\rangle\,\Theta\!\left[m-m_{\rm min}(M_{\rm WD})\right].
  \label{eq:rateMW}
\end{equation}
Figure~\ref{fig:rate} shows the result for $f_{\rm PBH}=1$. The rate peaks at $\Gamma_{\rm MW}\simeq 1.5$ per century for $m\simeq 3\times10^{21}\,{\rm g}$, and stays $\gtrsim 1$ per century for $2\times10^{21}\lesssim m\lesssim 2\times10^{22}\,{\rm g}$. It is cut off sharply at low mass by the ignition threshold and decreases as $1/m$ at high mass. 

Overall, the rate is comparable to the observed Galactic SN~Ia rate of a few per century~\cite{Korol2024}. Asteroid-mass PBHs can thus account for the entire SN~Ia rate without fine-tuning, with the rate scaling linearly with $f_{\rm PBH}$.

\end{document}